\documentclass[useAMS,usenatbib]{mn2e}
\usepackage{graphicx,lscape,amssymb}
\usepackage{natbib}
\usepackage{epstopdf}
\usepackage{fancyhdr}

\title{A photometric selection of White Dwarf candidates in SDSS DR10}

\author[Gentile Fusillo et al.]{Nicola Pietro Gentile Fusillo $^1$, Boris T. G\"ansicke $^1$, Sandra Greiss$^1$\\$^1$ Department of Physics, University of Warwick, Coventry, CV4 7AL, UK\\}

\begin{document}
\maketitle

\label{firstpage}

\begin{abstract}
We present a method which uses cuts in colour-colour and reduced proper motion-colour space to select white dwarfs without the recourse to spectroscopy while allowing an adjustable compromise between
completeness and efficiency. Rather than just producing a list of white dwarf candidates, our method calculates a \emph{probability of being a white dwarf} ($P_\mathrm{WD}$) for any object with available 
multi band photometry and proper motion. We applied this method to all objects in the SDSS DR10 photometric footprint and to a few selected sources in DR7 which did not have reliable 
photometry in DR9 or DR10. 
This application results in a sample of 61969 DR10 and 3799 DR7 photometric sources  with calculated $P_\mathrm{WD}$ from which it is possible to select a sample of $\sim23000$ high-fidelity white dwarf candidates with $T_\mathrm {eff} \gtrsim 7000$ K and $g\leq 19$. This sample contains over 14000 high confidence white dwarfs candidates which have not yet received spectroscopic follow-up. These numbers show that, to date, the spectroscopic coverage of white dwarfs in the SDSS photometric footprint is, on average, only $\sim40\%$ complete. While we describe here in detail the application of our selection to the SDSS catalogue, the same method could easily be applied to other multi colour, large area surveys.
We also publish a list of 8701 bright ($g\leq 19$) WDs with SDSS spectroscopy, of which 1781 are new identifications in DR9/10.
\end{abstract}


\section{Introduction}
White dwarfs (WD) are the stellar remnants left over from the evolution of stars with main sequence masses $M > 0.8M_\odot$ and $M\lesssim8-10 M_\odot$ (\citealt{ibenetal97-1},\citealt{smarttetal09-1}, \citealt{dohertyetal15-1}). This mass range includes over 90\% of all the stars in the Galaxy, making white dwarfs by far the most common stellar remnants.
Large samples of white dwarfs are required to reliably constrain fundamental parameters such as space density (\citealt{holbergetal02-1}, \citealt{holbergetal08-1}, \citealt{giammicheleetal12-1}, \citealt{sionetal14-1}), mass distribution (\citealt{bergeronetal92-1}, \citealt{liebertetal05-1},  \citealt{kepleretal07-1}, \citealt{falconetal10-1}, \citealt{tremblayetal13-1}, \citealt{Kleinmanetal13-1}) and luminosity function, 
which in turn can be used to determine the ages of the individual Galactic populations (\citealt{oswaltetal96-1}, \citealt{degennaroetal08-1}, \citealt{cojocaruetal14-1}).

Furthermore, well defined large samples of 
white dwarfs are an extremely useful starting point for identifying rare white dwarf types like magnetic white dwarfs (\citealt{gaensickeetal02-5}, \citealt{schmidtetal03-1}, \citealt{kuelebietal09-1}, \citealt{kepleretal13-1}), pulsating white dwarfs (\citealt{castanheiraetal04-1}, \citealt{greissetal14-1}), high/low mass white dwarfs (\citealt{vennes+kawka08-1}, \citealt{brownetal10-1}, \citealt{hermesetal14-1}), white dwarfs with unresolved low mass companions (\citealt{farihietal05-1}, \citealt{girvenetal11-1}, \citealt{steeleetal13-1}), white dwarfs with rare atmospheric composition (\citealt{schmidtetal99-1}, \citealt{dufouretal10-1}, \citealt{gaensickeetal10-1}), close white dwarf binaries (\citealt{marshetal04-1}, \citealt{parsonsetal11-1}), metal polluted white dwarfs (\citealt{sionetal90-1}, \citealt{zuckermanetal98-1}, \citealt{dufouretal07-2}, \citealt{koesteretal14-1}) or white dwarfs with dusty or gaseous planetary debris discs (\citealt{gaensickeetal06-3}, \citealt{farihietal09-1}, \citealt{debesetal11-2}, \citealt{wilsonetal14-1}).
Because of their intrinsic low luminosities identifying a large, complete and well defined sample of white dwarfs still remains a challenge. Much progress has been made in recent years thanks to large 
area surveys, first and foremost the Sloan Digital Sky Survey (SDSS, \citealt{yorketal00-1}) (\citealt{harrisetal03-1},  \citealt{eisensteinetal06-1}, \citealt{Kleinmanetal13-1}).
The largest published catalogue of white dwarfs to date (\citealt{Kleinmanetal13-1}) fully exploited the spectroscopic data available at the time of SDSS data release 7 and contains over 20000 white dwarfs (of which 7424 with $g\leq 19$).
However not only is DR7 now outdated, but SDSS spectroscopy is only available for less than 0.01\% of all SDSS photometric sources. Furthermore most of SDSS's white dwarfs are only serendipitous spectroscopic targets.
The true potential of SDSS's vast multi band photometric coverage still remains to be fully mined for white dwarf research, but this requires a reliable method able to select white dwarfs candidates without recourse to spectroscopy.
Proper motion has been traditionally used to distinguish white dwarfs from other stellar populations. In particular many studies that contributed to the  census of white dwarfs in the solar neighbourhood specifically targeted high proper motion objects (\citealt{holbergetal02-1}, \citealt{sayresetal12-1}, \citealt{limogesetal13-1}).
In this paper we present a novel method which makes use of photometric data and proper motions to calculate a \emph{probability of being a WD} ($P_\mathrm {WD}$) for any photometric source within a broad region in colour space. Unlike any previous similar work, our method does not use a specific cut in colour or proper motion to generate a list of white dwarf candidates; instead it provides a catalogue of sources with an associated $P_\mathrm {WD}$. These $P_\mathrm {WD}$ can then be used to create samples of white dwarf candidates best suited for different specific uses. By applying our method to the full photometric footprint of SDSS DR10, we created a catalogue which includes $\sim23000$ bright ($g\leq 19$) high-fidelity white dwarfs candidates.
Using this catalogue, we asses the spectroscopic completeness of the SDSS white dwarf sample.

\section{SDSS}
\label{sdss}
\begin{figure}
\includegraphics[width=\columnwidth]{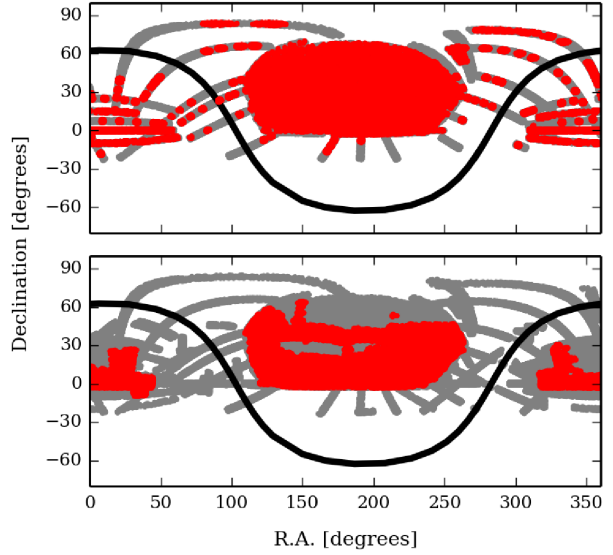}
\caption{\label{skycover}Photometric coverage (grey) of SDSS DR7 (top panel) and SDSS DR10 (bottom panel) in equatorial coordinates. The spectroscopic coverages of the SDSS-II spectrograph for DR7 and BOSS for DR10 are overlaid in red. The black line indicates the location of the galactic plane.}
\end{figure}

The Sloan Digital Sky Survey has been in continuous operation since 2000. It uses a dedicated 2.5 meter telescope at Apache Point in New Mexico to carry out multi band photometric observation of the northern sky and follow-up spectroscopy of selected targets. 
We have made use of the SDSS Data Release (DR) 7 (\citealt{abazajianetal09-1}), DR9 (\citealt{Ahnetal12-1}) and DR10 (\citealt{Ahnetal14-1}), which are, respectively, the last DR of the SDSS-II project and the second and third DRs of the SDSS-III project. 
All data releases provide $ugriz$ photometry spanning a magnitude range $\sim15-22$ and proper motions computed from the USNO-B and SDSS positions. 
In the sample  we examined as part of this work there is no object with a measured proper motion exactly equal to zero, but $\simeq 2\%$ of objects with magnitude $g \leq 19$ have no proper motion in the SDSS database. 
This is probably because these objects did not have a reliable match on the USNO-B photographic plates.
From here on we will refer to these objects as having no proper motion, even though their proper motions have, probably, simply not been computed and are actually not zero in value.  
SDSS DR7 includes photometric coverage of $11500\,\deg^2$ and follow-up low-resolution ($R\simeq1850-2200$, $3800-9200$\,\AA)
spectroscopy for 1.44 million galaxies, quasars, and stars.
In SDSS DR9 the photometric sky coverage was extended to a total $14555\,\deg^2$ which includes $2500\,\deg^2$ in the Southern Galactic Cap (Fig. \ref{skycover}).
In SDSS-III a new improved spectrograph called BOSS is used providing larger wavelength coverage ($3600-10400 \mathrm{\,\AA}$  \citealt{Ahnetal12-1}) as well as higher spectral resolution ($R\simeq1560-2650$). As of DR10 Sloan released over 3.35 million useful optical spectra.\\
Even though the main targets of BOSS spectroscopic follow-up are quasars and galaxies, about 3.5\% of the BOSS fibers in DR9 and DR10 were devoted to a series of 25 small ancillary projects.
Particularly relevant to our work is the white dwarf and hot subdwarf ancillary project which targeted $\sim5700$ white dwarf and hot subdwarf candidates selected according to
their  $u-r$, $u-g$, $g-r$ colours (\citealt{dawsonetal13-1}, \citealt{Ahnetal14-1}, Sect. 6.3).

\begin{table*}
\caption{\label{recap} Summary of the most relevant numbers presented in the paper}
\begin{tabular}{ll}
\hline
\hline
magnitude limit of the catalogue & $g\leq19$\\
Objects in main DR10 photometric catalogue (sect. 5) & 61969\\
Objects in DR7 extension (sect. 7.3) & 3799\\
Objects with DR7 spectra in initial colour cut & 28213\\
\hspace{1em} Poor quality spectra or no proper motion & 574\\
\hspace{1em} Objects in DR7 training sample (sect. 3, Table \ref{colour-cut}) & 27639\\
\hspace{1em} WDs in the DR7 training sample & 6706\\ 
\hspace{1em} Contaminants in the DR7 training sample &  20933\\
Object with SDSS/BOSS spectra in the catalogue & 33073\\
\hspace{1em} WDs with SDSS/BOSS spectra in the catalogue (Table \ref{DR9-spectra}) & 8701\\
High confidence WDs candidates in the catalogue & $\sim23000$\\
\hspace{1em}Of which with no SDSS spectra & $\sim14000$\\
WDs from \citet{Kleinmanetal13-1} included in our catalogue (sect. 8.1) & 6689\\
\hspace{1em}\citet{Kleinmanetal13-1} WDs not classifed as WDs by us & 30\\
\hspace{1em} Objects with a DR7 spectrum classified by us as WDs, not included in the  \citet{Kleinmanetal13-1} catalogue &  261\\
 & \\
\hline
\hline
\end{tabular}
\end{table*}

\section{Developing a photometric selection method}
\begin{table}
\caption{\label{e-cut} Equations describing the colour and magnitude 
constraints used to select primary sources in the SDSS footprint.}
\begin{tabular}{lcl}
\hline
Colour & constraint & \\
\hline
$(u-g)$ & $\le$ & $3.917 \times (g-r) + 2.344$\\
$(u-g)$ & $\le$ & $0.098 \times (g-r) + 0.721$\\
$(u-g)$ & $\ge$ & $1.299 \times (g-r) - 0.079$\\
$(g-r)$ & $\le$ & $0.450$\\
$(g-r)$ & $\ge$ & $ 2.191 \times (r-i)-0.638$\\
$(r-i)$ & $\le$ & $-0.452 \times (i-z) + 0.282$\\
$g$     & $\le$ & 19 \\
$type$  & $=$   & 6\\
\hline
\end{tabular}
\end{table}

\begin{figure*}
\includegraphics[width=2\columnwidth]{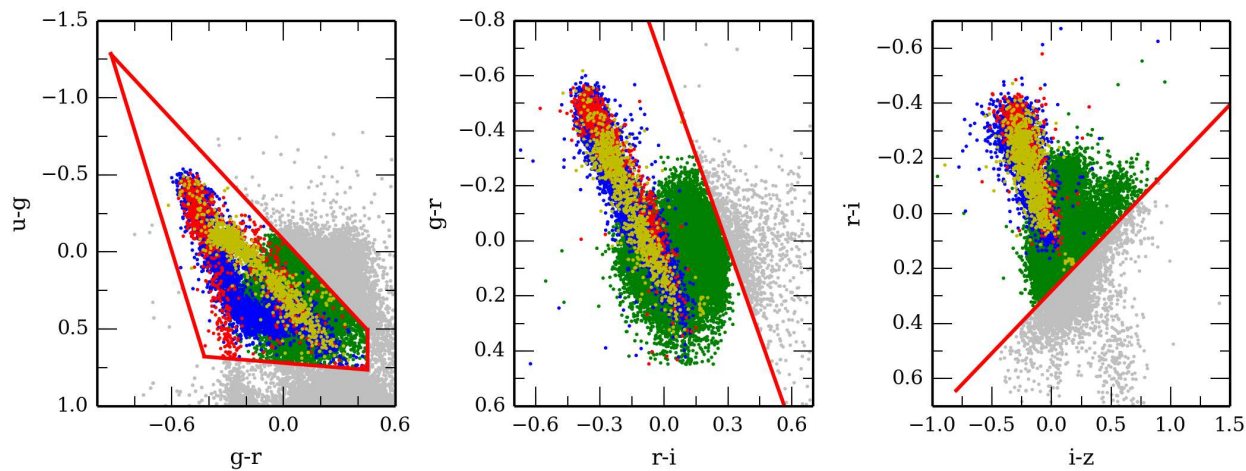}
\caption{\label{colour-cut} Colour-colour diagrams illustrating the location of
the 27639 DR7 spectroscopic objects that we used as training sample for our selection method. DA white dwarfs, non DA white dwarfs, NLHS and quasars are shown as blue, yellow, red and green
dots respectively. The colour cuts that define our initial broad selection from Table\,\ref{e-cut} are
overlaid as red lines. Objects outside this selection were not classified and are therefore plotted as grey dots.}
\end{figure*}

\begin{table}
\caption{\label{DR7-spectra} Classification of the 28213 objects with available spectra and with $g \leq19$ selected from DR7}
\begin{tabular}{lcl}
\hline
Class & number of objects\\
\hline
DA & $5271$ \\
DB & $497$ \\
DAB/DBA & $95$ \\
DAO & $49$ \\
DC & $404$ \\
DZ & $111$ \\
DQ & $120$ \\
Magnetic WD & $134$ \\
WD+MS & $197$ \\
CV & $94$ \\
NLHS & $1454$ \\
QSO & $19739$ \\
Unreliable & $36$ \\
Unclassified & $12$ \\
\hline
\end{tabular}
\end{table}

\begin{figure*}
\includegraphics[width=2\columnwidth ]{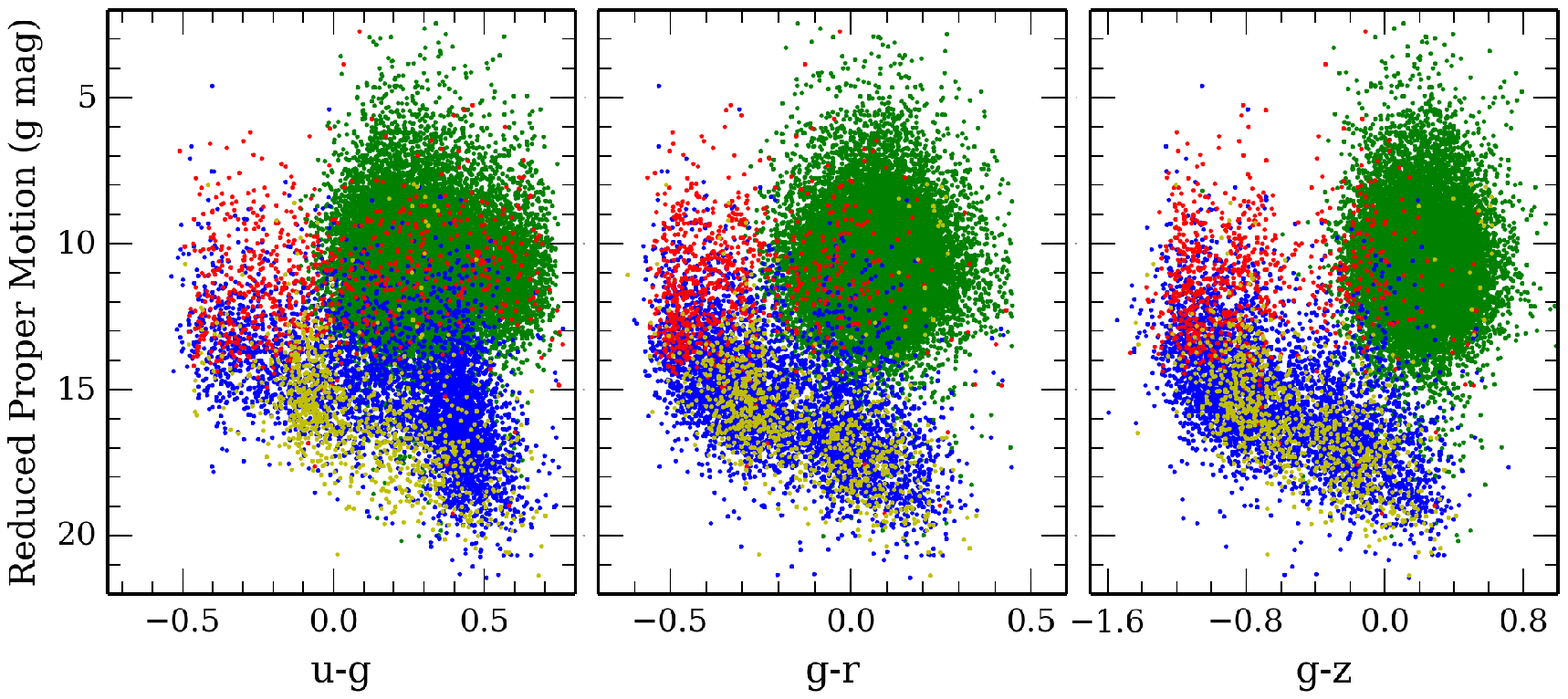}
\caption{\label{colour-rpm} Reduced proper motion-colour diagrams illustrating the location of
the 27639 DR7 spectroscopic objects of our training sample (Table\,\ref{DR7-spectra}). DA white dwarfs, non DA white dwarfs, NLHS and quasars are shown as blue, yellow, red and green
dots respectively.}
\end{figure*}
We first retrieved spectra, $ugriz$ photometry and proper motions for all the primary point sources with available spectra in DR7 within a broad region selected in the $(u-g,g-r)$, $(g-r,r-i)$, and
$(r-i,i-z)$ colour-colour planes (Fig.\,\ref{colour-cut}, Table\,\ref{e-cut}). 
The shape and extension of these colour-cuts were defined such that they included all of the objects which had been classified as either spectroscopically confirmed  white dwarfs or as photometric white dwarf candidates by \citet{girvenetal11-1}.
At this stage we were aiming to be as complete as possible and no real effort was made to avoid contamination.

In developing our selection method, we relied on visual classification of our initial spectroscopic sample and on proper motions. Sloan objects fainter than $g \sim 19$ often have noisy spectra and missing or unreliable proper motions. For this reason we decided to limit ourself to bright sources ($g \le 19$).

This first sample included 28213 objects which we classified according to spectral appearance. For the development of the selection method we only needed to classify these objects in 3 broad categories: "white dwarfs", "non white dwarfs" and "unreliable"(where the S/N was too low for classification). However we decided that a more detailed classification could help to diagnose biases during the development of the selection method and provide useful statistics. Therefore we subdivided the "white dwarfs" into 10 types (DA, DB, DC, Magnetic white dwarfs,... Table\,\ref{DR7-spectra}) and the "non WDs" into "QSOs" and a second category "Narrow Line Hydrogen Stars" (NLHS, a mixed bag of stars with low-gravity hydrogen dominated atmospheres, including subdwarfs, extreme horizontal branch stars and A/B type star). The NLHS sample may include a very small number of extremely low mass (ELM, \citealt{brownetal12-1}, \citealt{hermesetal14-1}, \citealt{gianninasetal14-1}) white dwarfs. However we correctly identified all but one  known ELM white dwarfs in our training sample (see Sect. 7.2 for a detailed discussion). The results of our classification are summarized in Table\,\ref{DR7-spectra}.\\
After discarding 36 objects with "unreliable" spectra, we calculated
reduced proper motions (RPMs) for all the objects in the sample, 

\begin{equation}H=g+5\log\mu+5\end{equation} 

\noindent with the Sloan $g$ magnitude and the proper motion $\mu$ in arcsec/year. 538 objects (of which 265 white dwarfs) did not have proper motions, reducing the size of our initial sample to 27639 (Table \ref{recap}).
These 27639 spectroscopically confirmed white dwarfs and contaminants with calculated RPM were the \textit{training sample} on which we developed our selection method.
RPM can be used as a proxy for absolute magnitude for a given transverse velocity and, with accurate photometry and astrometry, colour-RPM diagrams can show a very clean separation 
between main sequence stars, subdwarfs, white dwarfs and quasars.

The training sample was used to trace the loci occupied by white dwarfs and contaminants in RPM colour space and to explore the separation between the two types of objects achieved using different colours. We found that the strongest discrimination between white dwarfs and contaminants is obtained in $(g-z, \mathrm{RPM})$ space which we therefore adopted for our selection method (Fig. \ref{colour-rpm}).

We then mapped the distribution of the white dwarfs and contaminants of our training sample in RPM colour space. In order to create a smooth continuous map, every object was included as a 2D Gaussian the width of which reflects the uncertainty of the RPM and $(g-z)$ colour of the object. The Gaussians were normalised so 
that their volume equals unity and therefore the integral over the map is equal to the number of objects in the training sample. In this way we produced a continuous smeared-out "density map" for white dwarfs, and another one for contaminants.

The proper motions computed by SDSS for objects with  $g \leq 19$ are accurate to $\sim2.5$ mas/year \citep{Ahnetal12-1}, but many objects (most of the QSO) have proper motion values $<$ 2 mas/year and, consequently, large relative uncertainties. 
These values generate very large uncertainties in the computed RPMs and translate into Gaussians extremely stretched in the RPM dimension. These objects with poor proper motion measurements can, in fact, be smeared over the entire RPM dimension "polluting" even areas which should be populated only by the highest proper motion objects.
Furthermore by visually inspecting the $(g-z, \mathrm{RPM})$ distribution of QSO it is instantly obvious that they all cluster in a well defined 
locus which has a far smaller extension in the RPM dimension than the uncertainties of these low proper motions. 
To avoid such artifacts affecting our maps we decided to limit the maximum uncertainty in proper motion for any object to one third of the proper motion value. This correction only affects the objects with the lowest proper motions which, in the case of our training sample, are $\sim10000$ QSOs and  $\sim200$ other contaminants.

\begin{figure}
\includegraphics[width=\columnwidth]{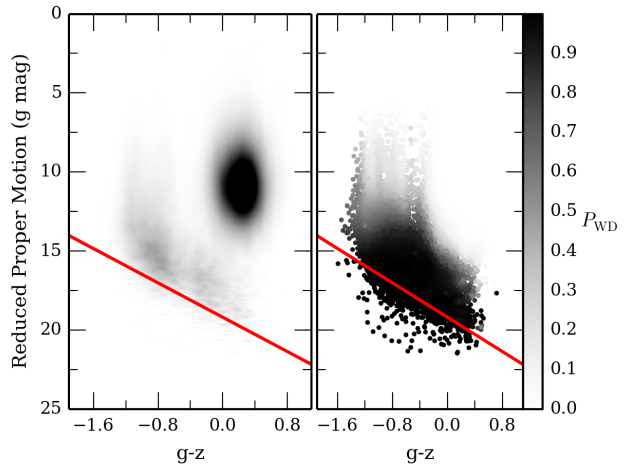}
\caption{\label{p-density} Distribution in $(g-z, \mathrm{RPM})$ of the 27639 white dwarfs and contaminants of the DR7 spectroscopic sample. In the left panel the objects are included as 2D Gaussians to account for the  uncertainties in their parameter 
and the gray-scale reflects the spatial density. In the right panel the grey-scale indicates 
the calculated $P_\mathrm {WD}$ with darker objects having higher values than lighter ones. All objects below the red line had their $P_\mathrm {WD}$ fixed to 1.0}
\end{figure}

We then defined a map providing the \textit{probability of being a white dwarf} ($P_\mathrm{WD}$), as the ratio of the white dwarf density map to  the sum of both density maps (white dwarfs and contaminants). 
$P_\mathrm{WD}$ of any given object is calculated by integrating the product of its Gaussian distribution in the $(g-z, \mathrm{RPM})$ plane with the underlying probability map (Fig.\,\ref{p-density}). For any given photometric source this value directly indicates how likely it is for the source to be a white dwarf.
Our  DR7 training sample only contained few objects with very high RPMs and therefore the probability map is scarcely populated in the regime of extremely high RPM. This leads to a "patchy" probability map with blank areas with no information. When calculating $P_\mathrm {WD}$ for objects outside our training sample, the blank areas, caused by lack of data, would generate low probability values which would not reflect any actual likelyhood of being a white dwarf.

To obviate this problem we decided to define a line in $(g-z, \mathrm{RPM})$ space such that the $P_\mathrm {WD}$ of all objects below this line
is assumed as 1.0. The line, given by
\begin{equation} \mathrm{RPM} > 2.72  \times (g-z)+19.19\end{equation}
\noindent was defined by inspecting the $(g-z, \mathrm{RPM})$ diagram of the spectroscopic sample and trying to include as much as the sparsely populated area as possible, 
while minimising the number of contaminants that would undergo such probability correction. (Fig.\,\ref{p-density}).

\begin{figure}
\includegraphics[width=\columnwidth]{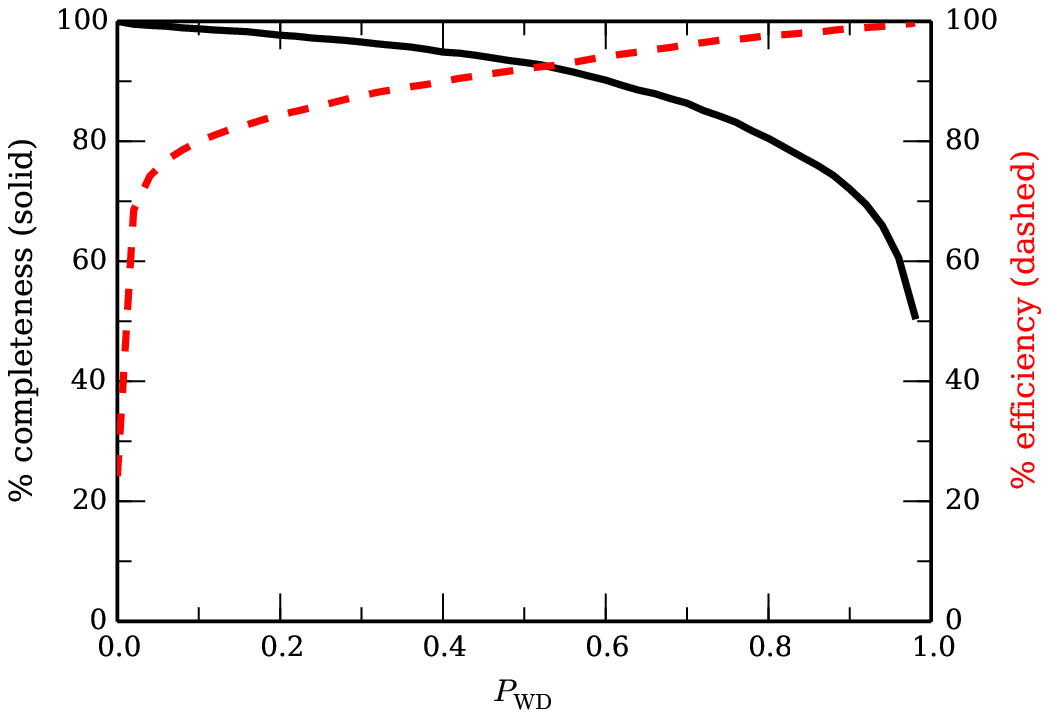}
\caption{\label{comp-eff}                                                                        Completeness (solid line) and efficiency (dashed line) of samples of white dwarf candidates from our catalogue shown as functions of the minimum value of $P_\mathrm {WD}$ an object must have in order to be selected. These values of completeness and efficiency were computed using the spectroscopic DR7 training sample as a reference.}
\end{figure}

\begin{figure}
\includegraphics[width=\columnwidth]{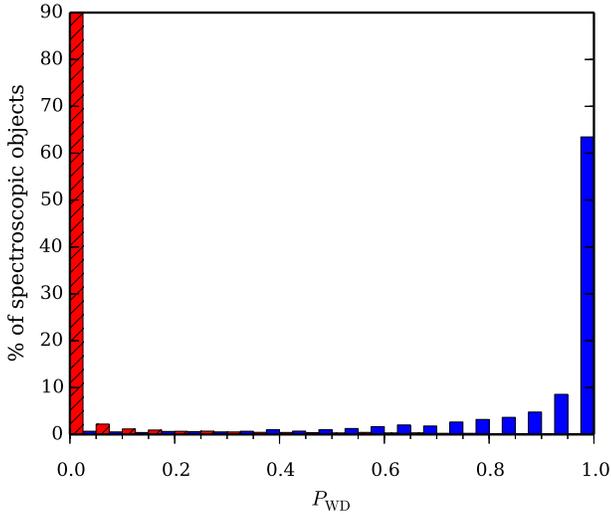}
\caption{\label{histo_DR7} Distribution of 27639 spectroscopically identified white dwarfs (blue) and contaminants (red, shaded) from the DR7 training sample as a function of $P_\mathrm {WD}$.}
\end{figure}

\begin{figure*}
\includegraphics[width=2\columnwidth ]{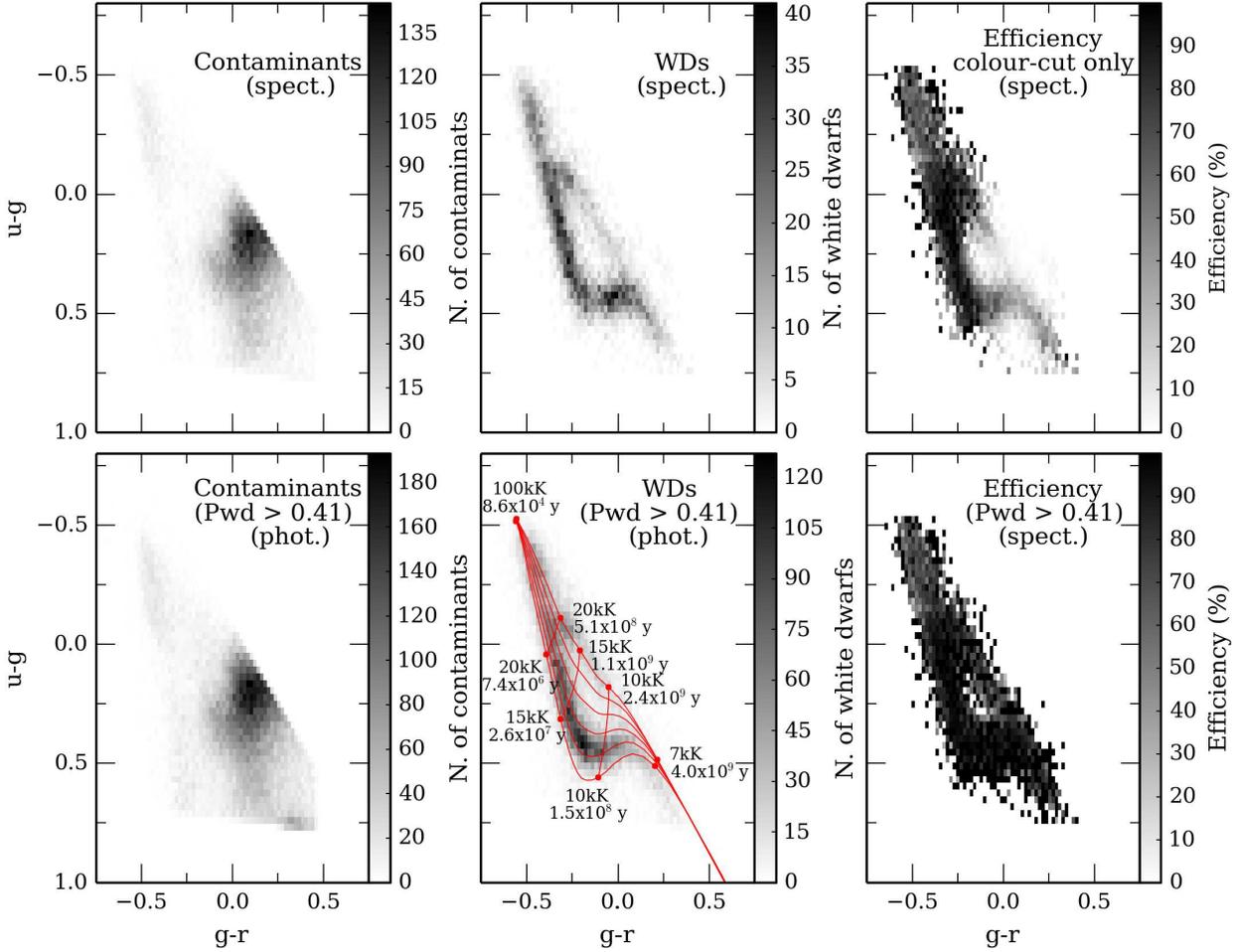}
\caption{\label{colour_cover} DR7 spectroscopic training sample (27639 objects, top) and DR10 photometric sample (61969 objects, bottom) within our initial $(u-g, g-r)$ colour-colour selection.
The top left and top middle panels show, respectively, the distribution of the contaminants and the white dwarfs from our initial DR7 spectroscopic sample. 
The top right panel shows the ratio of spectroscopically confirmed white dwarfs to the total number of objects with spectra in DR7,
$N_\mathrm{WD}/(N_\mathrm{WD} + N_\mathrm{Cont})$. The top right panel clearly illustrates the efficiency of a selection which only uses colour cuts. Such selection leaves areas of strong 
contamination at the red and the blue ends of the  $(u-g, g-r)$ colour region.
The bottom left panel shows the distribution of all sources from 
our DR10 photometric catalogue with $P_\mathrm {WD}$ $< 0.41$; these objects would be considered contaminants when compiling a 
95\% complete sample. 
Similarly, the bottom middle panel shows the distribution of all sources from 
our DR10 photometric catalogue with $P_\mathrm {WD}$ $\geq 0.41$, these objects would all be considered high-confidence white dwarf candidates when compiling 
a 95\% complete sample. White dwarf cooling tracks are shown in the red overlay. Both photometric distributions are extremely similar to their spectroscopically determined counterparts (top panels). The bottom right panel shows the ratio of spectroscopically confirmed white dwarfs with $P_\mathrm {WD}$ $\geq 0.41$ to the total number of objects with spectra in DR7 with $P_\mathrm {WD}$ $\geq 0.41$. This diagrams effectively shows the efficiency of our selection method. Unlike the top right panel, this probability based selection is practically independent of the location in colour space.}
\end{figure*}

Using the calculated $P_\mathrm {WD}$ it is now possible to make different confidence selections by defining any object
with an associated probability above an arbitrary threshold as a white dwarf candidate. When choosing such threshold value, completeness and efficiency are the key parameters one needs to compromise
between. Reference values of completeness and efficiency can be calculated using again our training sample. For a given $P_\mathrm {WD}$ threshold, we define \textit{completeness} as the ratio of the number of white dwarfs in the training sample with at least that associated probability to the total number of white dwarfs in the sample.
Similarly \textit{efficiency} is defined as the ratio of the number of white dwarfs selected by the probability cut to the number of all the objects retrieved by such selection.
While testing the selection method we determined that we can generate from our DR7 training sample a 95\% complete sample with 89.7\% efficiency by selecting objects with $P_\mathrm {WD}$ $\geq0.41$ (Fig.\,\ref{comp-eff}).
Fig.\,\ref{comp-eff} clearly shows that the efficiency of any confidence cut increases sharply up to probability values of $\sim0.08$ and more slowly after that. This effect is caused by the fact that the vast majority of the contaminants in our training sample are quasars with $P_\mathrm {WD} <0.1$ while most of the white dwarfs in the training sample have $P_\mathrm {WD} > 0.8$. Fig.\,\ref{histo_DR7} illustrates the distribution of white dwarfs and contaminants in terms of their $P_\mathrm {WD}$ and shows that there are indeed only a few objects with probabilities between 0.1 and 0.8. Therefore even if a confidence cut at probabilities of 0.1 already yields an efficiency of $\sim80\%$, much better completeness-efficiency compromises can be achieved using higher probability thresholds and $P_\mathrm {WD}$ so low should be chosen only when compiling a catalogue which aims to maximise completeness. 
Finally, in Fig.\ref{colour_cover} we use colour-colour diagrams to further illustrate the reliability of our selection method by comparing the a 95\% complete photometric sample with the DR7 spectroscopic training sample.

\section{White dwarfs with new spectra in DR9/10}
\begin{figure}
\includegraphics[width=\columnwidth]{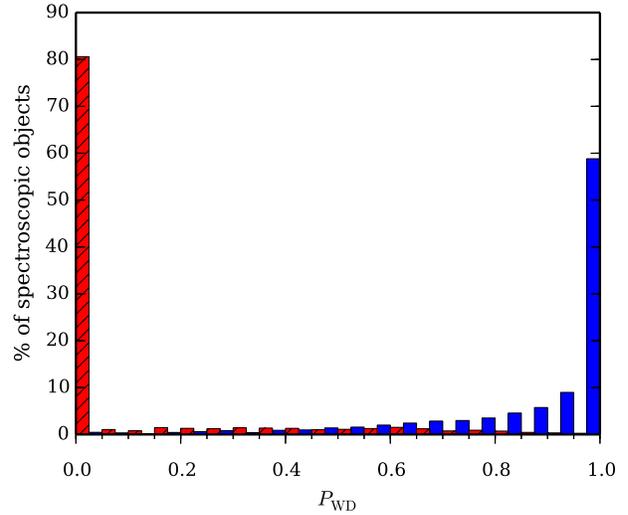}
\caption{\label{histo} Distribution of 8034 spectroscopically identified white dwarfs (blue) and contaminants (red, shaded) with proper motions from the sample of objects with new DR9/10 spectra} as a function of $P_\mathrm {WD}$.
\end{figure}

\begin{figure}
\includegraphics[width=\columnwidth]{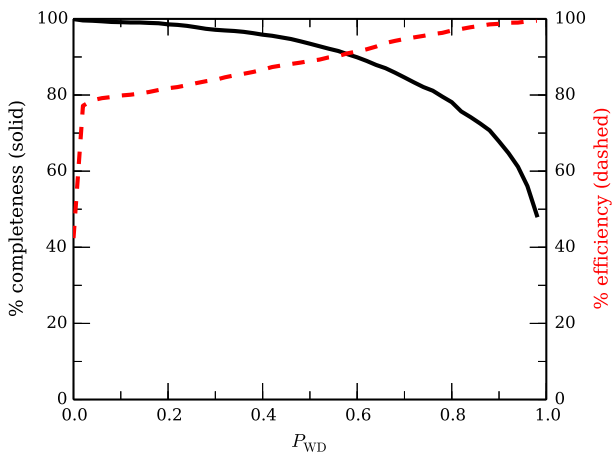}
\caption{\label{BOSS_comp-eff} 
Completeness (solid line) and efficiency (dashed line) of samples of white dwarf candidates from our catalogue shown as functions of the minimum value of $P_\mathrm {WD}$ an object must have in order to be selected. These values of completeness and efficiency were computed using the SDSS-III spectroscopic sample as a reference.}
\end{figure}

Using again the  broad selection described in Table \,\ref{colour-cut}, we retrieved $ugriz$ photometry, proper motions and spectra for $8215$ objects which received spectroscopic follow-up after DR7 up as part of SDSS-III. These are predominately objects observed with the BOSS spectrograph, including only 102 targets of the Segue-II program still observed with the old SDSS spectrograph. We classified the spectra by visual inspection (Table\,\ref{DR9-spectra}).
The 3560 objects identified as white dwarfs form an independent sample of spectroscopically confirmed white dwarfs which we used to further test the reliability of our selection method. Furthermore 1752 of these white dwarfs did not have a spectrum prior to DR9 and are therefore new spectroscopically confirmed white dwarfs. We used the $(g-z, \mathrm{RPM})$ 
probability map to estimate $P_\mathrm {WD}$ for 3522 of these 3560 white dwarfs (since 38 of them did not have proper motion) and verify that their $P_\mathrm {WD}$ computed from the DR7 probability map are consistent with the spectroscopic classification.
Fig.\,\ref{histo} clearly shows that the vast majority of the white dwarfs with new DR9/10 spectra have $P_\mathrm {WD}$ $>$ 0.6 and over 80\% of spectroscopically confirmed contaminants have $P_\mathrm {WD}< 0.1$  confirming that our selection method can reliably distinguish between white dwarfs and contaminants.
As a further test, we also decided to calculate values of completeness and efficiency using only objects with new DR9/10 spectra (Fig.\,\ref{BOSS_comp-eff}) in the same way we did before using the DR7 training sample (Sect. 3). 
Even though this new spectroscopic sample is considerably smaller than the DR7 training sample, the calculated values of completeness and 
efficiency are similar; e.g. selecting objects with $P_\mathrm {WD}\geq0.41$ achieves a completeness of 96\% and an efficiency of 86.4\% on the DR9/10 spectroscopic sample, comparable to the completeness of 95\% and an efficiency of 89.7\% achieved for the training sample.

\begin{table}
\caption{\label{DR9-spectra} Classification of the 8215 objects with spectra taken after DR7, with $g \leq19$ within the initial broad colour-cut.
The new spectroscopically confirmed white dwarfs had not received any spectroscopic follow up before DR9.}
\begin{tabular}{lcl}
\hline
Class & number of objects\\
\hline
DA & $2488$ \\
DB & $408$ \\
DAB/DBA & $127$ \\
DAO & $46$ \\
DC & $214$ \\
DZ & $44$ \\     
DQ & $57$ \\
CV & $27$ \\
Magnetic WD & $60$ \\
WD+MS & $89$ \\
NLHS & $902$ \\
QSO & $3735$ \\
Unreliable & $16$ \\
Unclassified & $2$ \\
\hline
New spectroscopically & $1752$\\  
confirmed WDs & \\
\hline
\end{tabular}
\end{table}
\section{A catalogue of photometric white dwarfs candidates in DR10}
We retrieved $ugriz$ photometry and proper motions for all the primary point sources in DR10 using the criteria described in Table\,\ref{e-cut}, but adding the additional constraint 
that the selected objects must have  proper motions. Furthermore, having to rely only on photometric data we decided to also exclude objects that were flagged as having too few good detections and saturated pixels. This results in a total of 61969 photometric objects. We calculated 
RPMs for these objects and, using the probability map created with the DR7 training sample (Sect. 3), we calculated their $P_\mathrm {WD}$.
Our catalogue can be easily used as the starting point for creating different white dwarf candidates samples according to the compromise between completeness and efficiency best suited for different specific uses.
Table\,\ref{Col_tab} illustrate the structure and the content of the catalogue, the full list of objects can be accessed online via the VizieR catalogue access tool.

\begin{table*}
\centering
\caption{\label{Col_tab} Format of the DR10 catalogue of white dwarf candidates. The full catalogue can be accessed online via the VizieR catalogue tool. }
\begin{tabular}{lll}
\hline
\hline
Column No. & Heading & Description\\
\hline
1 & sdss name & SDSS objects name (SDSS + J2000 coordinates)\\
2 & SDSS-III photoID & Unique ID identifing the photometric source in SDSS-III\\
3 & SDSS-II photoID & Unique ID identifing the photometric source in SDSS-II\\
4 & ra & right ascension (J2000)\\
5 & dec & Declination (J2000)\\
6 & ppmra & proper motion in right ascension (mas/yr)\\
7 & ppmra err & proper motion in right ascension uncertainty (mas/yr)\\
8 & ppmdec & proper motion in right declination (mas/yr)\\
9 & ppmdec err & proper motion in right declination uncertainty (mas/yr)\\
10 & probability & The \emph{probability of being a WD} computed for this object\\
11 & umag & SDSS $u$ band PSF magnitude\\
12 & umag err & SDSS $u$ band PSF magnitude uncertainty\\
13 & gmag & SDSS $g$ band PSF magnitude\\
14 & gmag err & SDSS $g$ band PSF magnitude uncertainty\\
15 & rmag & SDSS $r$ band PSF magnitude\\
16 & rmag err & SDSS $r$ band PSF magnitude uncertainty\\
17 & imag & SDSS $i$ band PSF magnitude\\
18 & imag err & SDSS $i$ band PSF magnitude uncertainty\\
19 & zmag & SDSS $z$ band PSF magnitude\\
20 & zmag err & SDSS $z$ band PSF magnitude uncertainty\\
21 & instrument & instrument used to take the most recent spectrum of the object (SDSS or BOSS)\\
22 & specobjID SDSS-III & unique ID identifying the spectroscopic source in SDSS-III\\
23 & specobjID SDSS-II & unique ID identifying the spectroscopic source in SDSS-II \\
24 & human class & classification of the object based on our visual inspection of its spectrum\\
25 & Kleinman class & classification of the object according to \citet{Kleinmanetal13-1}\\
26 & Kepler class & classification of the object according to \citet{kepleretal15-1}\\
27 & ancillary flag & 1 indicates the object was part of the BOSS WD and subdwarfs ancillary program\\
28 & SDSS WD & 1 indicates that the objects was classified as a WD based on available SDSS spectra\\
29 & BOSS WD & 1 indicates that the objects was classified as a WD based on available BOSS spectra\\
30 & Brown WD flag & 1 indicates that the objects was classified as a WD in the the hypervelocity stars\\ 
   &   &  spectroscopic survey (\citealt{brownetal06-1}, \citeyear{brownetal07-1}, \citeyear{brownetal07-2}). 2 indicates that the object was\\ 
   &   &  classified as something other than a WD\\
31 & Simbad classification & Currently available Simbad classifications\\
32 & DR7 extension & 1 indicates that the objects was included as part of the DR7 extension (Sect. 7.3)\\
\hline
\end{tabular}

\end{table*}

\section{SDSS spectroscopic coverage}
\subsection{SDSS objects with multiple spectra}
About $24\%$ of the objects in the spectroscopic samples inspected as part of this work have multiple spectra resulting from repeat observations of plates or overlapping regions between plates. Most of these have 2-4 spectra taken with either SDSS or BOSS, but we also found a few white dwarfs with up to 16 spectra.
Concerning the work we describe in this paper, multiple spectra were only inspected for objects with a dubious classification. However, these spectra are a precious resource which can be used to investigate systematic uncertainties in stellar parameter obtained by spectral modelling and to probe for variability of spectral features.
For these reasons, in addition to our photometric catalogue of white dwarf candidates, we also provide a list of the available spectra (identifiable via \textit{MJD, plate ID} and \textit{fiber ID}) for all the objects in our catalogue (including the DR7 extension, sect. 7.3).

\begin{figure}
\includegraphics[width=\columnwidth]{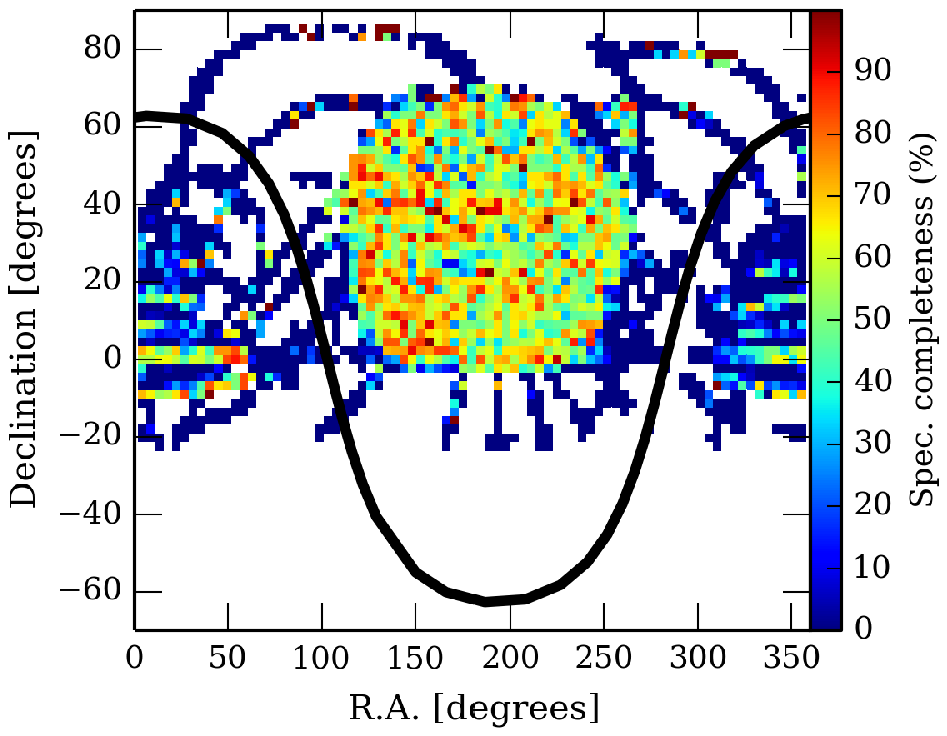}
\caption{\label{sky_bin} 
Spectroscopic completeness of SDSS white dwarfs satisfying the criteria in Table \ref{e-cut}, computed as the ratio of spectroscopically confirmed white dwarfs to all high-confidence white dwarf candidates ($P_\mathrm {WD}$ $\geq 0.41$) over the entire photometric footprint of SDSS (Table \ref{Col_tab}. The black line indicates the location of the galactic plane.}
\end{figure}

\begin{figure*}
\includegraphics[width=\columnwidth]{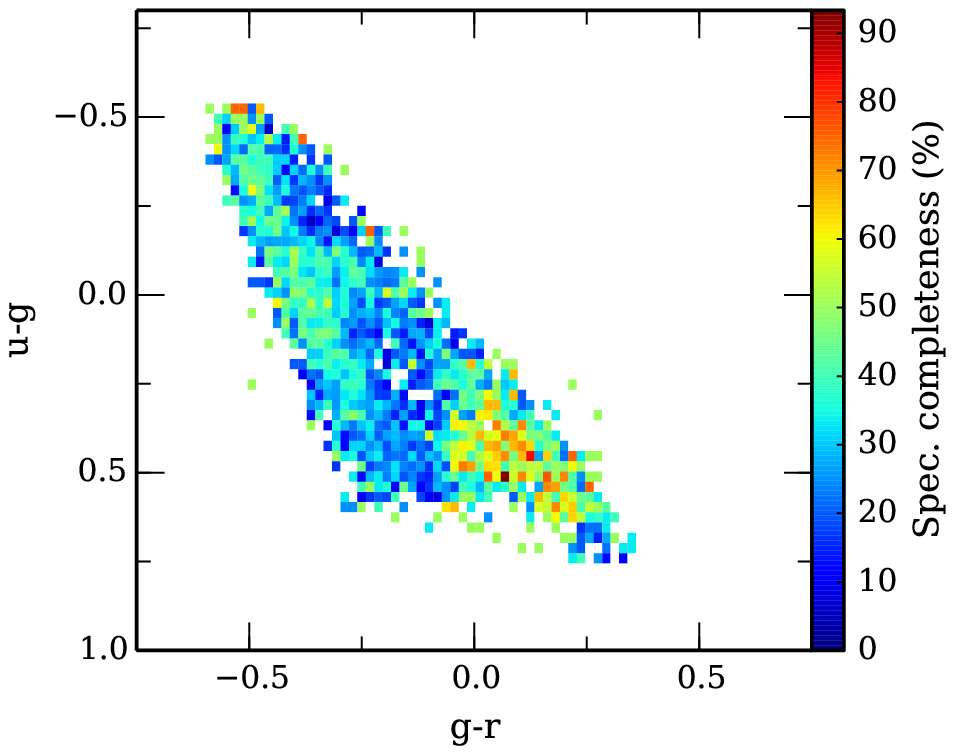}
\includegraphics[width=\columnwidth]{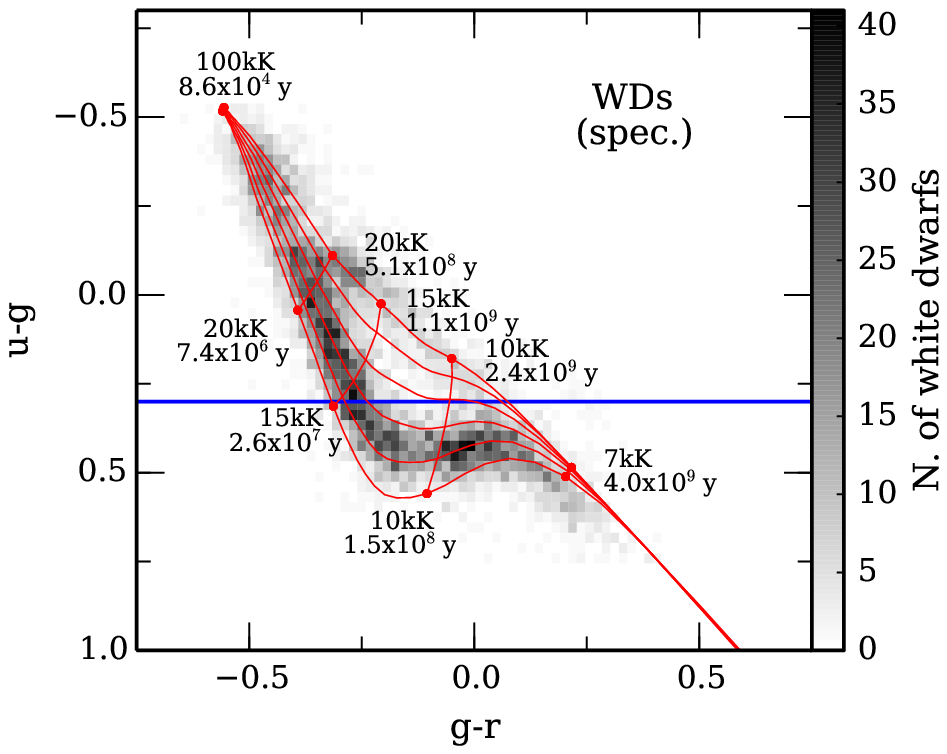}
\caption{\label{SDSS_comp} \emph{Left panel}: Spectroscopic completeness of SDSS white dwarfs, computed as the ratio of spectroscopically confirmed white dwarfs to all high-confidence white dwarf candidates ($P_\mathrm {WD}$ $\geq 0.41$) within our initial $(u-g,g-r)$ colour-colour selection.
\newline \emph{Right panel}: Distribution of spectroscopically confirmed white dwarfs within our initial $(u-g,g-r)$ colour-colour selection with cooling tracks shown as overlay. The blue line indicates the $u-g$ cut applied by the SDSS-III white dwarf and hot subdwarf stars ancillary project target selection. Only objects above the blue line were targeted, excluding most of the cool white dwarfs.}
\end{figure*}

\subsection{SDSS white dwarf spectroscopic completeness}
Using the $P_\mathrm {WD}$ in our catalogue and correcting for the completeness of the sample (Fig. \ref{comp-eff}) one can reliably estimate the total number of bright ($g \leq 19$) white dwarfs with $T_\mathrm {eff} \gtrsim 7000$ K in the SDSS photometric footprint. 
The ratio of spectroscopically confirmed white dwarfs to this total number of white dwarfs can be used as an estimate of the spectroscopic completeness of SDSS white dwarfs both in terms of spatial and colour distribution.
As mentioned before, most of the SDSS white dwarfs are only serendipitous spectroscopic targets, so it comes with no surprise that the average spectroscopic completeness of SDSS white dwarfs is only $\sim$40\%. However, this number is averaged over the entire SDSS photometric footprint, large areas of which have not yet received any spectroscopic follow up (Fig. \ref{skycover}). In Fig. \ref{sky_bin} we show that over the spectroscopic footprint (which covers most of the northern galactic cap), the average spectroscopic completeness of SDSS white dwarfs is actually closer to $\sim$75\%.

\begin{figure}	
\includegraphics[width=\columnwidth]{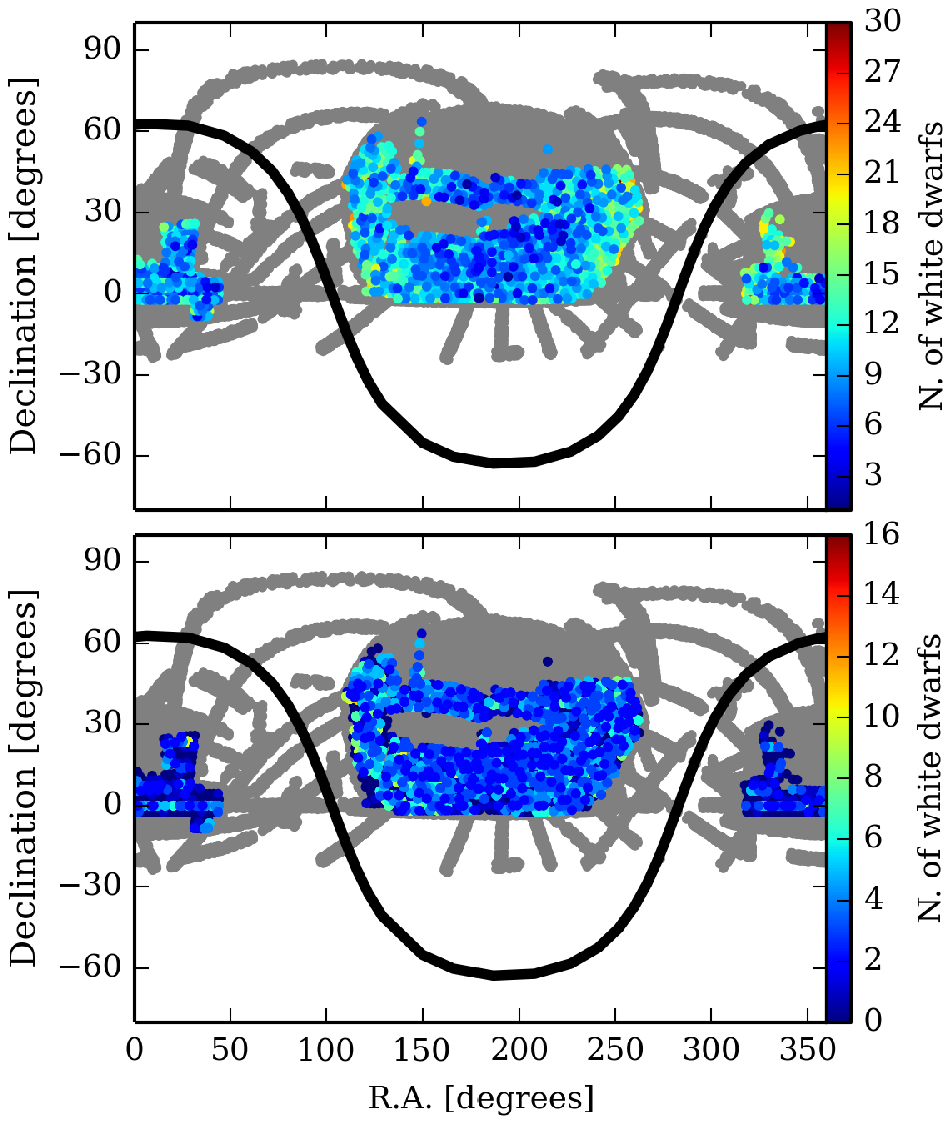}
\caption{\label{SDSS_colour_comp} Location of BOSS plates overlaid on the DR10 photometric footprint (grey). The black line indicates the location of the galactic plane.
\newline \emph{Top panel}: The colour of the plates indicates the number of high confidence white dwarf candidates ($g \leq 19$) from our catalogue per plate.
\newline \emph{Bottom panel}: The colour of the plates indicates the number of spectroscopically confirmed bright ($g \leq 19$) white dwarfs which were observed on the BOSS plate and released in DR9/10.
\newline Both panels clearly illustrate that the number of white dwarfs (or white dwarf candidates) per plate decreases with increasing distance from the galactic plane }
\end{figure}

However, as shown in Fig. \ref{SDSS_comp}, the spectroscopic completeness is also very colour-dependent.
Because quasars have always been one of the main targets of SDSS, the colour region populated by quasars has received more intense spectroscopic follow up. Consequently, the spectroscopic completeness of SDSS white dwarfs is highest for white dwarfs with colours similar to those of quasars and therefore $T_\mathrm {eff} \lesssim 10000$ K. At this cool end in $(u-g,g-r)$ colour space the spectroscopic completeness can be as high as $85\%$. On the other hand, the colour space occupied by hotter white dwarfs ($T_\mathrm {eff} \gtrsim 10000$ K) has received much sparser spectroscopic follow-up, which is reflected by the drop in spectroscopic completeness which varies between  $\sim$20\% and $\sim$40\%. Even though these less complete colour regions have been specifically covered by SDSS's ancillary white dwarf follow-up programs  \citep{dawsonetal13-1}, the number of white dwarfs observed by these programs only marginally affects the overall completeness (Sect. 6.3).
Fig. \ref{sky_bin} and \ref{SDSS_comp} clearly illustrate that the current sample of white dwarfs with Sloan's spectroscopy is inhomogeneous both in sky and colour distribution and extreme care should be taken when using it to compute any statistics.
Each BOSS plates covers an area of $1.49$ $\mathrm{deg}$ radius and has 1000 fibers.
However, on average, only $\sim 4$ white dwarfs were targeted on each BOSS plate, with only 2-3 plates having up to  $\sim 20$ white dwarfs and over 350 plates with no white dwarfs at all. Using our photometric catalogue of white dwarf candidates we estimated that on average $\sim13$ white dwarfs could have  been targeted on each BOSS plate, with some more densely white dwarf-populated plates at low galactic latitudes (Fig. \ref{SDSS_colour_comp}).
Even these very simple estimates already show that SDSS, or any other similar multi-object spectroscopic survey (i.e. LAMOST \citealt{zhaoetal13-1}, \citealt{zhangetal13-1}; WEAVE \citep{weave14-1}; 4MOST \citealt{4most14-1}), could easily provide spectroscopic follow-up of almost all bright white dwarf candidates with very little expenditure of fibers. In fact, a BOSS-like survey could produce a $>95\%$ complete, magnitude limited ($g \leq 19$), spectroscopic sample of white dwarfs  by dedicating just over 1\% of its fibers to white dwarf follow up. With such a complete and well defined spectroscopic sample it would be possible to carry out extremely reliable and diverse statistical analyses and finally answer many of the open questions about the formation and evolution of white dwarfs and their progenitors. Furthermore such a large spectroscopic sample would include many rare types of white dwarfs.

\subsection{SDSS-III white dwarf and hot subdwarf stars ancillary project}

About 3.5\% of the BOSS fibers in DR9 and DR10 were devoted to 25 small ancillary programs. 
One of these ancillary programs, the SDSS-III white dwarf 
and hot subdwarf stars ancillary project, specifically targeted 5709 white dwarf and hot subdwarf candidates selected using colour and proper motion as summarized in Table \ref{ancill-cut} (Dawson et al. 2013, Ahn et al. 2014).
Using the corresponding ancillary target flag \citep{dawsonetal13-1} we retrieved spectra, $ugriz$ photometry and proper motions of all the targets of this ancillary program. 
\begin{table}
\centering
\caption{\label{ancill-cut} Constraints used to select the targets of the
SDSS-III white dwarf and hot subdwarf stars ancillary project \citep{dawsonetal13-1}.}
\begin{tabular}{lcl}
Only areas of  DR7 footprint & $<$ & 0.5 mag\\
with galactic extinction in $r$ &  \\
\hline
Colour & constraint & \\
\hline
$(u-g)$ & $<$ & $0.3$\\
$(g-r)$ & $<$ & $0.5$\\
$(g-r)$ & $>$ & $-1$\\
$(u-r)$ & $<$ & $0.4$\\
$g$     & $<$ & 19.2 \\
\hline
For objects with & & \\
\hline
$(g-r)$ & $>$ & $-0.1$\\
$(u-r)$ & $>$ & $-0.1$\\
\hline
proper motion & $>$ & 20 mas/yr\\
\hline
\hline
\end{tabular}
\end{table}
4104 of these SDSS ancillary targets match the criteria (in either DR7 or DR10) in Table\,\ref{e-cut} and have proper motions and, therefore, must have been visually classified by us in either the DR7 training sample or the DR9/10 spectroscopic sample.
When comparing our classification to the targeting selection of the ancillary project we established that 78\% of the ancillary project targets included in our catalogue were classified as white dwarfs by us (Table\,\ref{ancil_comp}), with the vast majority of the rest being subdwarfs (e.g. \citealt{reindletal14-1}). 
This comparison shows that the targeting strategy adopted in this ancillary program was very  efficient, however the criteria in Table\,\ref{ancill-cut} limit the selection to white dwarfs with hydrogen dominated atmospheres (DA) hotter than $\sim14000$ K and white dwarfs with helium dominated atmospheres (DB) hotter than $\sim8000$ K (Table \ref{ancill-cut}).
In conclusion the SDSS-III white dwarf and hot subdwarf stars ancillary project significantly  contributed in increasing the number of SDSS white dwarfs with spectra, but this sample of white dwarfs produced suffers from various biases and should be handled with extreme care when used for statistical analysis. 

\begin{table}
\centering
\caption{\label{ancil_comp} Results of the  comparison of the SDSS-III white dwarf and hot subdwarf stars ancillary project with our classification of SDSS spectra.
\newline After inspecting  the spectra of the 904 ancillary targets not classified as white dwarfs by us, we are confident that these are most likely hot subdwarfs.}

\begin{tabular}{ll}
\hline
\hline
Total number of WDs from the ancillary & 4104\\
program included in our catalogue & \\
including DR7 extension &\\
\hline
Number of ancillary program targets not & 904\\
classifed as WDs. & \\

\hline
\hline
\end{tabular}
\end{table}

\section{Limitations and corrections}
 
\subsection{Proper motions}
One of the main limitation of the selection method used to generate this catalogue is that our $P_\mathrm {WD}$ can only be calculated for objects with proper motions. 3.8\% of all the spectroscopically confirmed white dwarfs we examined do not have proper motion and we can therefore expect that our DR10 photometric catalogue is incomplete by $\sim870$ white dwarfs candidates on account of them not having proper motion.
\subsection{Extremely Low Mass white dwarfs (ELM white dwarfs)}
Even though our spectroscopic training sample, initially drawn from a broad colour selection (Fig. \ref{colour-cut}), includes the vast majority of single white dwarfs (see Sect.8) it does not include some rarer types of white dwarfs with more exotic colours. Extremely low mass (ELM) white dwarfs are likely to be among such rare types \citep{brownetal10-1}. We tested the ability of our selection method to correctly identify ELM white dwarfs by retrieving the SDSS phototmetry of all bright ($g \leq 19$) ELM white dwarfs from \citet{gianninasetal14-1} 
and verifying whether or not they were recovered in our catalogue. We determined that only 17 out of 37 are included in our catalogue. One of the ELM white dwarfs was excluded because of bad photometry in SDSS, but the remaining 19 were simply outside our initial colour cut. 
Furthermore ELM white dwarfs have peculiar spectra which can make it hard to correctly classify them as white dwarfs. Of the 17 ELM white dwarfs in our catalogue 11 have either SDSS or BOSS spectra and were therefore visually classified by us. Only one of these ELM white dwarfs was erroneously classified as a NLHS. Consequently we believe that miss-classification of ELM white dwarfs in our training sample does not significantly affect our selection method. In fact only 2 of the ELM white dwarfs in our catalogue have $P_\mathrm {WD}$ $< 0.41$ while the remaining 16 have $P_\mathrm {WD}$ $> 0.6$.
We conclude that ELM white dwarfs within our colour regions would most likely be identified as high confidence white dwarfs candidates.

\subsection{DR7 to DR10 miss-matches: the "DR7 extension"}
When trying to retrieve the 6706 white dwarfs with proper motions of our DR7 spectroscopic training sample (Sect. 2) from our DR10 photometric catalogue, we noticed that some DR7 objects could not be correctly matched in DR10.
The cross matching of DR7 spectra to DR10 photometric sources was done using modified Julian date (MJD), plate ID and fiber ID which, together, uniquely identify any
spectrum in SDSS. This was faster and more reliable than a coordinate cross match using the entire DR10 photometric database or a specifically selected subset of it (some objects underwent changes in coordinates, photometry or flags in between data releases). Furthermore, in this way, we were able to check that the DR7 spectra were still available in DR10 and to check for other possible incongruences between the data releases.  
To our surprise we found that our DR10 photometric sample only included 6560 of the spectroscopically confirmed white dwarfs.
We determined that 11 of the "missing objects" could not be matched to sources in DR10 because the MJD with which their spectra were identified changed between DR7 and DR10 \footnote{From the spectra we examined, we determined that MJDs uniquely identifying SDSS spectra have changed between DR7 and 10 for the following plates: 0389, 2129, 2516, 2713, 2865.}. However the objects are included 
in the DR10 photometric sample and therefore the cross-match was manually corrected and our count was updated to 6571 retrieved DR7 white dwarfs. 
The remaining 135 white dwarfs are missing from the photometric DR10 sample because of differences in their photometry or flags in DR10 compared to DR7 (Table\,\ref{ext_reason}). 

\begin{table}
\centering
\caption{\label{ext_reason} Detailed break down of the reasons why 135 white dwarfs from our DR7 training sample do not figure in the main DR10 photometric catalogue.
\newline 29 objects marked as "colours changed" experienced changes in their recorded $ugriz$ magnitudes such that in DR10 they would not fall into our initial colour cut.
\newline 23 objects marked as "Not in SDSS-III database" either fall in the SEGUE-1 imaging scans which were not included in SDSS-III or are in areas of the sky where, probably because of the new SDSS-III sky subtraction, faint sources near a bright source were not identified \citep{Ahnetal12-1}.}

\begin{tabular}{ll}
\hline
N. of WDs "missing" & Cause\\
\hline
17 &  $type \neq 6$ in DR10\\
63 & $gmag > 19$ in DR10\\
29 & colours changed\\
3  & No recorded proper motion in DR10\\
23 & Not in SDSS-III database\\
\hline
135 & \\
\hline
\end{tabular}
\end{table}

In order to make our catalogue as complete as possible we decided to add a ``DR7 extension'' to our original list of 61969 DR10 white dwarf candidates. This extension was created by recovering $ugriz$ photometry for all reliable primary photometric 
sources in DR7 that satisfied the criteria described in Table\,\ref{e-cut} and then selecting only those objects that did not appear among the 61969 DR10 photometric candidates. The cross-matching between DR9/DR10 and DR7 photometric 
sources was done by comparing sky coordinates using a 1 arcsecond matching radius which generated a list of 3799 DR7 objects. 689 of these do not have any counterpart in DR10 as they either fall in the SEGUE-1 imaging scans which were not included in SDSS-III or are in areas of the sky where, probably because of the new SDSS-III sky subtraction, faint sources near a bright source were not extracted \citep{Ahnetal12-1}. The remaining 3110 are objects whose 
photometric data or flags have changed between data releases such that they no longer fulfil the criteria in Table\,\ref{e-cut} in DR9/10.
As a final check we confirmed that the ``DR7 extension'' includes all of the 135 white dwarfs that were spectroscopically identified in DR7, but which were missing in our DR10 photometric catalogue (Table\,\ref{ext_reason}). Furthermore within the "DR7 extension" we were also able to identify and classify 29 new white dwarfs which only have BOSS spectroscopy (Table\,\ref{BOSS_indr7ext}).

\begin{table}
\caption{\label{BOSS_indr7ext} Classification of the 297 objects with BOSS spectra, found in our "DR7 extension". 29 of the new spectroscopically confirmed white dwarfs had not received spectroscopic follow-up before DR9. }
\begin{tabular}{lcl}
\hline
Class & number of objects\\
\hline
DA & $32$ \\
DB & $7$ \\
DAB/DBA & $1$ \\
DAO & $0$ \\
DC & $4$ \\
DZ & $0$ \\     
DQ & $1$ \\
Magnetic WD & $0$ \\
WD+MS & $6$ \\
CV & $2$ \\
NLHS & $21$ \\
QSO & $222$ \\
Unreliable & $1$ \\
Unclassified & $0$ \\
\hline
New spectroscopically & $29$\\  
confirmed WDs & \\
\hline
\end{tabular}
\end{table}

\subsection{DR10 to DR7 miss-matches}

In the previous paragraph we discussed how a few objects which fulfilled our selection criteria (Table. \ref{e-cut}) in DR7 did not do so in DR10.
Similarly we expected to find some sources with DR7 spectra which did not fulfil our selection criteria in DR7 but do so in DR10 and therefore would be included in our main photometric sample.
From our 61969 photometric candidates we selected all objects with SDSS-II spectroscopy which were not included in our initial DR7 spectroscopic sample, and therefore already classified.
We identified 2071 of these SDSS-II spectroscopic sources which we had not yet inspected even though they were included in of our catalogue.

Once again the main reasons for the "appearance" of these sources in our DR10 selection are: changes in their $g$ magnitude which may have moved an object below our  $g \le 19$ limit; corrections of flags of objects previously, erroneously, classified as "extended sources" and inclusion of proper motions for objects which did not previously have one.
However, we also identified 31 sources in our DR10 sample with SDSS spectra which simply do not have any DR7 photometry (in our catalogue they lack a DR7 photometric ID).
In order to keep our catalogue as consistent and complete as possible, we decided to retrieve   spectra for these 2071 sources (using the DR7 photometric ID included in our table) and proceeded to visually classify them.
The results of the classification of these 2071 objects are shown in Table \ref{addDR7-spectra}.

\begin{table}
\caption{\label{addDR7-spectra} Classification of the 2071 objects with SDSS spectra which did not fulfil our selection criteria in DR7 but did so in DR10}
\begin{tabular}{lcl}
\hline
Class & number of objects\\
\hline
CV & $6$ \\
DA & $136$ \\
DB & $12$ \\
DAB & $27$ \\
DAO & $1$ \\
DC & $14$ \\
DZ & $3$ \\     
DQ & $5$ \\
Magnetic WD & $1$ \\
WD+MS & $9$ \\
NLHS & $95$ \\
QSO & $1675$ \\
Unclassified & $2$ \\
unreliable & $84$ \\
\hline
\end{tabular}
\end{table}

\section{Comparison with other catalogues}{\label{other_cat}}
\subsection{Kleinman et al 2013}
The \citet{Kleinmanetal13-1} catalogue of spectroscopically identified DR7 white dwarfs contains 20407 spectra corresponding to 19712 unique objects of which only 7424 are brighter than $g$=$19$. 
Unlike our spectroscopic samples (DR7 training sample and DR9/DR10 BOSS sample) the \citet{Kleinmanetal13-1} catalogue does not have  a set magnitude limit and includes spectra of extremely faint objects. Most of these have low signal to noise ratios making the classification, for at least some of them, inevitably less reliable.
In order to carry out a comparison between our spectral classification and Kleinman's we proceeded to cross match the \citet{Kleinmanetal13-1} catalogue of DR7 white dwarfs with the DR10 \textit{specphotoall} 
table using modified Julian date (MJD), plate ID and fiber ID and obtained SDSS-III IDs for these objects. Analogously to what is described in section 6.2, 52 spectra could initially not be matched 
to DR10 sources because of changes in the MJD and another 149 spectra could not be matched because the corresponding photometric sources are not included in the DR10 photometric database.
861 more white dwarfs were rejected by our selection because of their colour, flags or lack of recorded proper motion (Table\,\ref{klei_reason}). 
After manually correcting the miss-matching MJDs and including the "DR7 extension" discussed before, we established that our catalogue includes 6689 objects classified as white dwarfs by \citet{Kleinmanetal13-1}.

\begin{table}
\caption{\label{klei_reason}Detailed break down of the reasons why 861 white dwarfs from the  \citet{Kleinmanetal13-1} catalogue are not included in our main DR10 photometric catalogue.
Most of these are white dwarfs with a main sequence companion. 
Note that 119 of these were recovered in our "DR7 extension" (sect 7.3)}
\label{klei_reason}
\begin{tabular}{lc}
\hline
Keinman WDs excluded in & Cause\\
DR10 photometric catalogue &  \\
\hline
601 & Excluded by criteria\\
 &  in Table\,\ref{e-cut}\\
260 & No proper motion\\
\hline
861 & \\
\hline
\end{tabular}
\end{table}

Inspecting the results of our spectroscopic classification, we find that we also identified 99.6\% of these 6689 objects as white dwarfs.  Only 30 (0.4\%) of Kleinman's white dwarfs were classified as contaminants by us when preparing the spectroscopic training sample (Sect. 3). We closely re-examined all available spectra for these 30 objects. We confirm our initial classification  of 20 objects as contaminants (18 NLHS and 2 QSO), and one object as "unclassified" because of the very low quality of its spectrum. However we concede that the remaining 9 objects are most likely white dwarfs and our initial classification was wrong.

Our catalogue also includes 261 objects with DR7 spectroscopy which we classified as white dwarfs, but which are not included in \citet{Kleinmanetal13-1} (Table\,\ref{recap}). \citet{Kleinmanetal13-1} do \textit{not} include a list of all spectra that they processed, and we have therefore no way to assess whether they inspected those objects. We conclude that our classification is remarkably close to Kleinman's and the disagreement over a very limited number of objects in our training sample does not effect our selection method.

Furthermore, unlike \citet{Kleinmanetal13-1}, we aim to provide a well defined sample SDSS white dwarfs candidates without being limited by the availability of spectroscopy. Our catalogue contains $\sim23000$ high-confidence bright ($ g \leq19$) white dwarfs candidates, over 3 times the number of white dwarfs in  \citet{Kleinmanetal13-1} catalogue with the same magnitude limit, underlining the remaining  potential for follow up spectroscopy.

\subsection{Hypervelocity stars. I,II,III (Brown et al 2006, 2007a, 2007b)}{\label{hyper}}
The hypervelocity stars spectroscopic survey (\citealt{brownetal06-1}, \citeyear{brownetal07-1}, \citeyear{brownetal07-2}) identified a total of 260 white dwarfs serendipitously found among 
B star candidates. 48 of these white dwarfs have  $g >19$ and are therefore not included in our photometric sample; another 9 are excluded by our initial selection criteria (Table\,\ref{e-cut}) and 23 more do not have proper motions in SDSS DR9/10. 
This leaves 180 white dwarfs spectroscopically identified by Brown et al. which are included in our DR10 photometric catalogue and which we flag as "known white dwarfs". Our catalogue also includes  296 hypervelocity survey targets which were classified as "non white dwarfs" by Brown et al. We flagged these objects to recognise them as "known contaminants".
Even though this sample of confirmed white dwarfs and contaminants is quite small, it was selected and classified completely independently from our work. Therefore these white dwarfs and contaminants are valuable test objects to verify once more the reliability of our $P_\mathrm {WD}$.

We established that the vast majority of the 180 white dwarfs have 
$P_\mathrm {WD}$ greater than 0.9 and only 11 of them have probabilities lower than 0.6. 
Out of the 296 contaminants 34 have $P_\mathrm {WD}$ greater than 0.5, however only 11 
of them have probability values higher than 0.7.
In conclusion our $P_\mathrm {WD}$ seems to be somewhat weaker when trying to separate these contaminants as they were specifically selected as objects with colours similar to those of typical white dwarf \textit{and} to have high proper motions.

\subsection{Follow-up spectroscopy: 11 new white dwarfs}
\begin{figure*}
\vspace*{0.5cm}
\hspace*{0.5cm}
\includegraphics[width=1.9\columnwidth ]{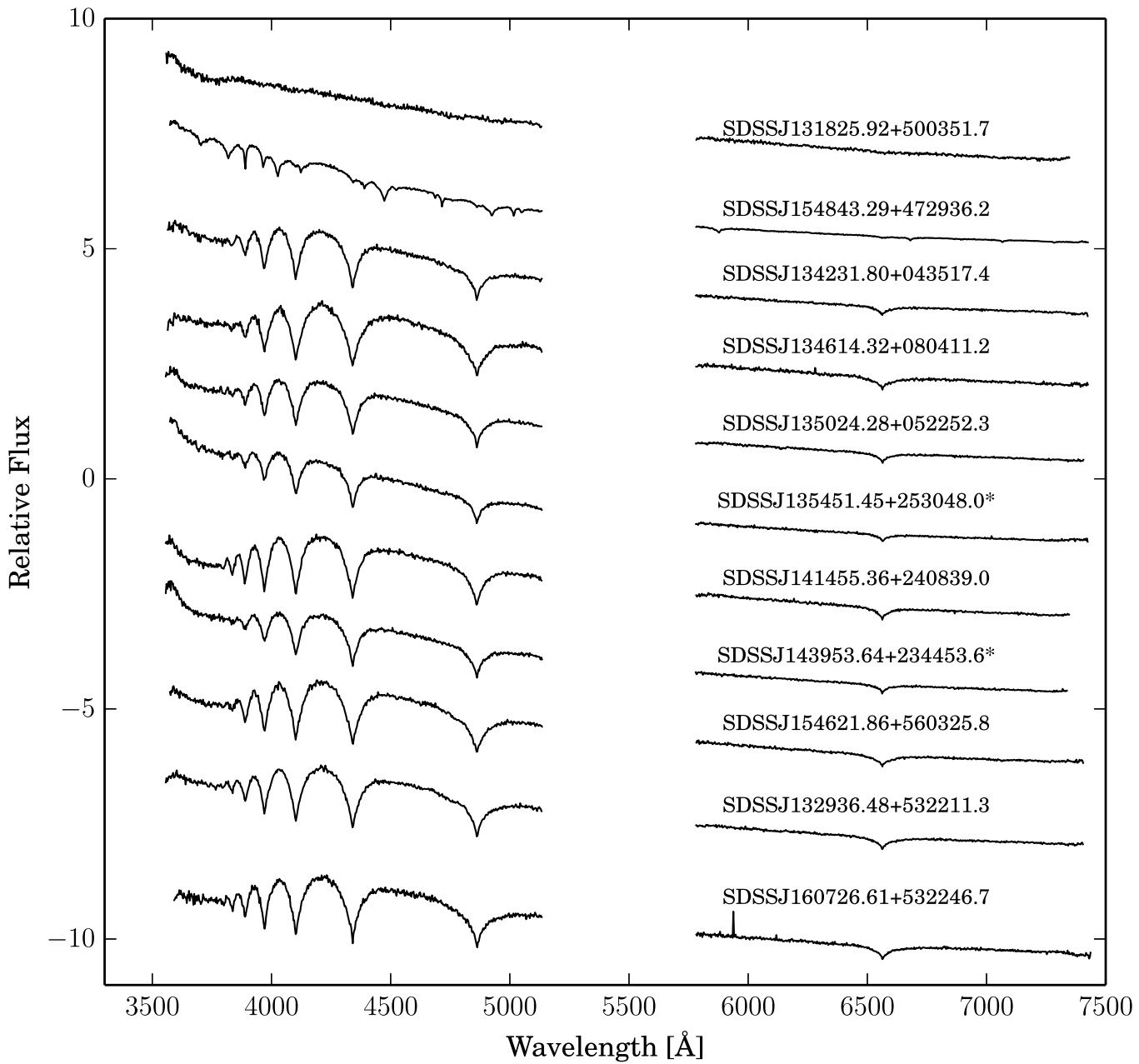}
\vspace{1.2cm}
\caption{\label{isisWD} WHT ISIS spectra (blue arm + red arm; R600 grating) of the 11 white dwarf candidates which were confirmed as white dwarfs. Each spectrum is labelled with the SDSS name. Objects marked with * have a new BOSS spectrum released in DR10}
\end{figure*}

\begin{figure*}
\vspace*{1cm}
\hspace*{1cm}
\includegraphics[width=1.9\columnwidth ]{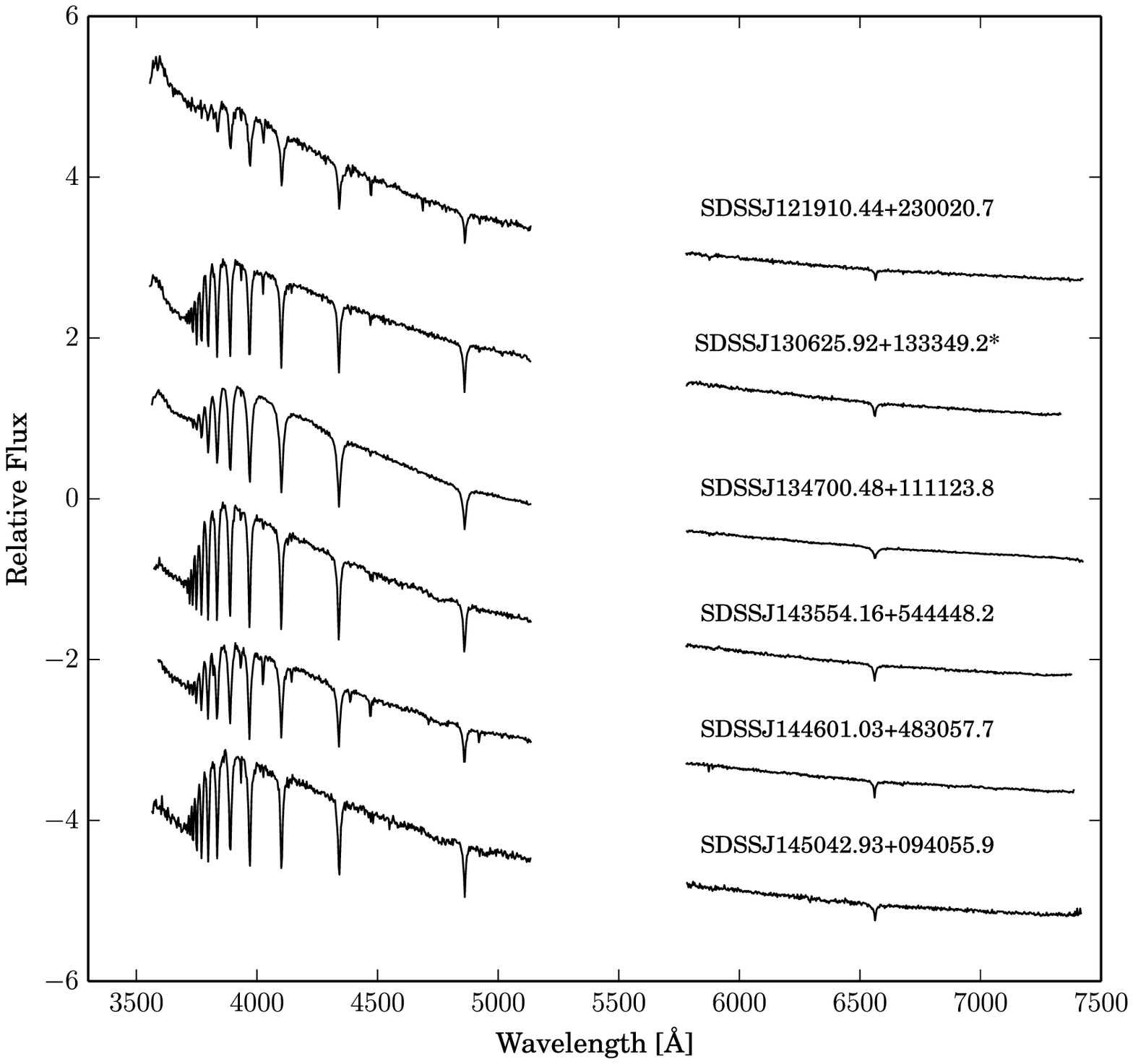}
\vspace{1.2cm}
\caption{\label{isisnnWD} WHT ISIS spectra (blue arm + red arm; R600 grating) of the 6 white dwarf candidates which were later classified as NLHS. Each spectrum is labelled with the SDSS name. Objects marked with * have a new BOSS spectrum released in DR10}
\end{figure*}

As a final test for our selection method we selected 17 white dwarfs candidates from our catalogue which did not have a SDSS or BOSS spectrum (at the time of DR9) and spanned a wide range of $P_\mathrm {WD}$. 
On June 7$^{th}$ and 8$^{th}$ 2013, we obtained spectroscopy of these 17 objects using the double-armed Intermediate Resolution Spectrograph\footnote{http://www.ing.iac.es/Astronomy/instruments/isis/} (ISIS) on the William Herschel Telescope (WHT) on the island of La Palma. We observed under 1" seeing conditions. We used the R600R and R600B gratings in the ISIS blue and red arms respectively, with a 1" slit. 
The blue arm was centred at 4351$\mathrm{\AA}$ and the red arm at 6562$\mathrm{\AA}$. The blue spectra covered a total wavelength range from $\sim$3700$\mathrm{\AA}$ to $\sim$5000$\mathrm{\AA}$ and the red spectra ranged from $\sim$5700$\mathrm{\AA}$ to $\sim$7200$\mathrm{\AA}$. The spectral resolution is $\sim$2\,${\rm \AA}$ and $\sim$1.8\,${\rm \AA}$ in the red and in the blue arm respectively. Two consecutive 10 minute exposures were taken in order to increase the signal-to-noise ratio (S/N) of the average spectrum.
We adopted a standard reduction and calibration for the spectra \citep{greissetal14-1}. 
The spectra allowed us to classify our candidates and corroborate the validity of our $P_\mathrm {WD}$: the 9 targets with $P_\mathrm {WD}$ $\geq 0.89$ were all confirmed  as white dwarfs, 
out of 4 intermediate probability candidates  ($>0.45, <0.7$) only the highest probability candidate was confirmed as a white dwarf
and finally only one of the 4 low probability targets ($<0.39$) is a white dwarfs (Table\,\ref{ISIS}, Fig.\ref{isisWD}, Fig.\ref{isisnnWD}).
The release of DR10 added new BOSS spectra, compared to DR9, and 3 of the photometric candidates we spectroscopically followed-up now also have BOSS spectra (SDSSJ1354+2530, SDSSJ1439+2344, SDSSJ1306+1333; marked with * in Table.\,\ref{ISIS}). Visual inspection of these BOSS spectra confirmed our classification based on the ISIS spectra. 

\begin{table*}
\caption{\label{ISIS} Results of classification of the newly acquired ISIS spectra for 17 white dwarfs candidates from our catalogue. In the case of DA white dwarfs we also include effective temperature and surface gravity calculated by fitting the spectra with 1-D atmospheric models.}
\begin{tabular}{*{8}{c}}
\hline
SDSS name & probability & classification & $T_\mathrm {eff}$ (K) & log g\\
          & of being a WD & & &     \\
\hline
SDSSJ141455.36+240839.0 &  $1.000$ & $DA$ & 15071 $\pm$ 72 & 7.560 $\pm$ 0.018\\
SDSSJ134231.80+043517.4 &  $0.999$ & $DA$ & 20566 $\pm$ 65 & 7.770 $\pm$ 0.019\\
SDSSJ135024.28+052252.3 &  $0.999$ & $DA$ & 22037 $\pm$ 94 & 8.010 $\pm$ 0.020\\
SDSSJ143953.64+234453.6* & $0.993$ & $DA$ & 24726 $\pm$ 9  & 8.080 $\pm$ 0.010\\
SDSSJ160726.61+532246.7 &  $0.991$ & $DA$ & 12976 $\pm$ 98 & 8.010 $\pm$ 0.030\\
SDSSJ131825.92+500351.7 &  $0.985$ & $DC$ & - & -\\
SDSSJ154621.86+560325.8 &  $0.941$ & $DA$ & 15422 $\pm$ 103 & 7.950 $\pm$ 0.017\\
SDSSJ135451.45+253048.0* & $0.933$ & $DA$ & 28389 $\pm$ 54 & 7.960 $\pm$ 0.024\\
SDSSJ132936.48+532211.3 &  $0.891$ & $DA$ & 17707 $\pm$ 68 & 7.860 $\pm$ 0.016\\
SDSSJ134614.32+080411.2 &  $0.696$ & $DA$ & 16525 $\pm$ 91 & 8.260 $\pm$ 0.022\\
SDSSJ145042.93+094055.9 &  $0.566$ & $NLHS$ & - & -\\
SDSSJ121910.44+230020.7 &  $0.523$ & $NLHS$ & - & -\\
SDSSJ144601.03+483057.7 &  $0.455$ & $NLHS$ & - & -\\
SDSSJ130625.92+133349.2* & $0.355$ & $NLHS$ & - & -\\
SDSSJ154843.29+472936.2 &  $0.315$ & $DB$ & - & -\\
SDSSJ143554.16+544448.2 &  $0.288$ & $NLHS$ & - & -\\
SDSSJ134700.48+111123.8 &  $0.250$ & $NLHS$ & - & -\\
\hline
\end{tabular}
\end{table*}

\section{Conclusion}
We developed  a selection method for white dwarfs which allows to reliably select white dwarf candidates relying on their colours and reduced proper motion. By using the distribution of a large sample of spectroscopically confirmed white dwarfs and contaminants in colour-RPM space we calculate   a \emph{probability of being a WD} ($P_\mathrm {WD}$) for any object with available multi-band photometry and proper motion.\\
The spectroscopic sample used to develop our selection method was 
created by classifying over 27000 spectra of blue objects from SDSS DR7. In developing our selection method we decided limit our efforts to bright objects ($g\leq 19$) as fainter sources are less likely to have reliable detections on the historic photographic plates, necessary to calculate proper motions.
We used our selection method to calculate $P_\mathrm {WD}$ for 61969 primary photometric point sources selected from SDSS DR10 and for 3799 primary point sources selected from SDSS DR7 which fulfilled our selection criteria in DR7, but have incomplete data in the following DRs. This catalogue of over 65000 objects with calculated $P_\mathrm {WD}$ contains 8699 objects with available Sloan spectroscopy which we classified as white dwarfs. Using different values of $P_\mathrm {WD}$ as threshold, one can produce white dwarf candidates samples based on the compromise between completeness and efficiency best suited for different specific uses.
Even though we developed our method using SDSS DR7 and here we present its application to SDSS DR10, it can be used on any sample of objects having multi band photometry and proper motions.
We estimate that our SDSS DR10 photometric catalogue contains  $\sim23000$ high confidence white dwarf candidates of which $\sim14000$ have not been followed up spectroscopically.
These statistics imply that the spectroscopic sample of SDSS white dwarfs is currently, on average, only $\sim40\%$ complete for white dwarfs with $T_\mathrm {eff} \gtrsim 7000$ K and $g\leq 19$, underlining the remaining potential for follow-up spectroscopy.  

\section{acknowledgements}
NPGF acknowledge the support of Science
and Technology Facilities Council (STFC) studentships.
\newline The research leading to these results has received funding from the
European Research Council under the European Union's Seventh Framework
Programme (FP/2007-2013) / ERC Grant Agreement n. 320964 (WDTracer).
\newline Funding for SDSS-III has been provided by the Alfred P. Sloan Foundation, the Participating Institutions, the National Science Foundation, and the U.S. Department of Energy Office of Science. The SDSS-III web site is http://www.sdss3.org/.
We thank S.O. Kepler for a constructive referee report and Daniel Eisenstein for comments on the originally submitted manuscript.
The William Herschel Telescope is operated on the island of La Palma by the Isaac Newton Group in the Spanish Observatorio del Roque de los Muchachos of the Instituto de Astrofísica de Canarias.

\appendix
\section{Comparison with Kepler et al. 2015}
After submission of this paper, \citet{kepleretal15-1} 
posted a catalogue of 9088 white dwarfs and subdwarfs
spectroscopically identified among the SDSS
spectra obtained after DR7. For completeness,
we compare our classification of objects with new DR9/10 spectra (Sect. 4) with Kepler's catalogue.
Our catalogue includes 2041 white dwarfs and 590 subdwarfs  from the \citet{kepleretal15-1} catalogue.
Inspecting the results of our spectroscopic classification, we find that of these 2041 objects classified as white dwarfs by Kepler, 88.5\% were also identified as white dwarfs by us.  234 (11.5\%) of Kepler's white dwarfs were, instead, classified as contaminants by us (Sect. 4).
We closely re-examined all available spectra for these 234 objects, and confirm our
classification of NLHS for 223 objects, but concede that the remaining 11 are likely white
dwarfs. From this comparison with \citet{kepleretal15-1} and also by inspecting \citet{werneretal14-1}, we conclude that  most of the white dwarfs that we erroneously
classified as NLHS are DAO, PG1159 and O(He) (pre-) white dwarfs. Given that these
are rare objects, their erroneous classification does not affect the statistical properties of our selection method and white dwarf candidate sample".

Our catalogue also includes 57 objects with DR9/10 spectroscopy which we classified as white dwarfs, but which are not included in Kepler's list (Table\,\ref{kepler_comp}). Kepler do not include a list of all spectra that they processed, and we have therefore no way to assess whether they inspected those objects.

\begin{table}
\caption{\label{klei_reason} Result of the comparison between our classification of objects with spectra taken after DR7 and \citet{kepleretal15-1} catalague.}
\label{kepler_comp}
\begin{tabular}{lc}
\hline
WDs from \citet{kepleretal15-1} included & 2041\\
in our catalogue & \\
subdwarfs from \citet{kepleretal15-1} included &  590\\
in our catalogue & \\
\citet{kepleretal15-1} WDs not classified as WDs by us & 234\\
\citet{kepleretal15-1} subdwarfs classified as WDs by us & 5\\
Objects with a DR9/10 spectrum classified by us as & 57\\
WDs, not included in the \citet{kepleretal15-1} catalogue & \\
\hline
\end{tabular}
\end{table}
\bibliographystyle{mn_new}

\end{document}